\documentclass[
twocolumn,
]{ceurart}

\usepackage[frozencache,cachedir=.]{minted}
\setminted{breaklines=true}

\usepackage{graphicx}
\usepackage{subcaption}
\usepackage{lmodern}
\begin{document}

%
\copyrightyear{2024}
\copyrightclause{Copyright for this paper by its authors.
  Use permitted under Creative Commons License Attribution 4.0
  International (CC BY 4.0).}

\conference{}

\title{Beyond Following: Mixing Active Initiative into Computational Creativity}


\author[1]{Zhiyu Lin}[%
email=zhiyulin@gatech.edu,
url=https://zhiyulin.info/,
]
\author[1]{Upol Ehsan}[%
email=ehsanu@gatech.edu,
url=https://www.upolehsan.com/,
]
\author[1]{Rohan Agarwal}[%
email=rohanagarwal@gatech.edu,
]
\author[1]{Samihan Dani}[%
email=sdani30@gatech.edu,
]
\author[1]{Vidushi Vashishth}[%
email=vvashishth3@gatech.edu,
]
\author[1]{Mark Riedl}[%
email=riedl@cc.gatech.edu,
url=https://eilab.gatech.edu/mark-riedl.html,
]

\address[1]{Georgia Institute of Technology,
Atlanta, Georgia, USA
}




\begin{abstract}
Generative Artificial Intelligence (AI) encounters limitations in efficiency and fairness within the realm of Procedural Content Generation (PCG) when human creators solely drive and bear responsibility for the generative process.
Alternative setups, such as Mixed-Initiative Co-Creative (MI-CC) systems, exhibited their promise. 
Still, the potential of an active mixed initiative, where AI takes a role beyond following, is understudied.
This work investigates the influence of the adaptive ability of an active and learning AI agent on creators' expectancy of creative responsibilities in an MI-CC setting. 
We built and studied a system that employs reinforcement learning (RL) methods to learn the creative responsibility preferences of a human user during online interactions.
Situated in story co-creation, we develop a Multi-armed-bandit agent that learns from the human creator, updates its collaborative decision-making belief, and switches between its capabilities during an MI-CC experience.
With 39 participants joining a human subject study,
Our developed system's learning capabilities are well recognized compared to the non-learning ablation, corresponding to a significant increase in overall satisfaction with the MI-CC experience.  
These findings indicate a robust association between effective MI-CC collaborative interactions, particularly the implementation of proactive AI initiatives, and deepened understanding among all participants.
\end{abstract}


\maketitle

\section{Introduction}

\begin{figure*}
    \centering
    \includegraphics[width=\textwidth]{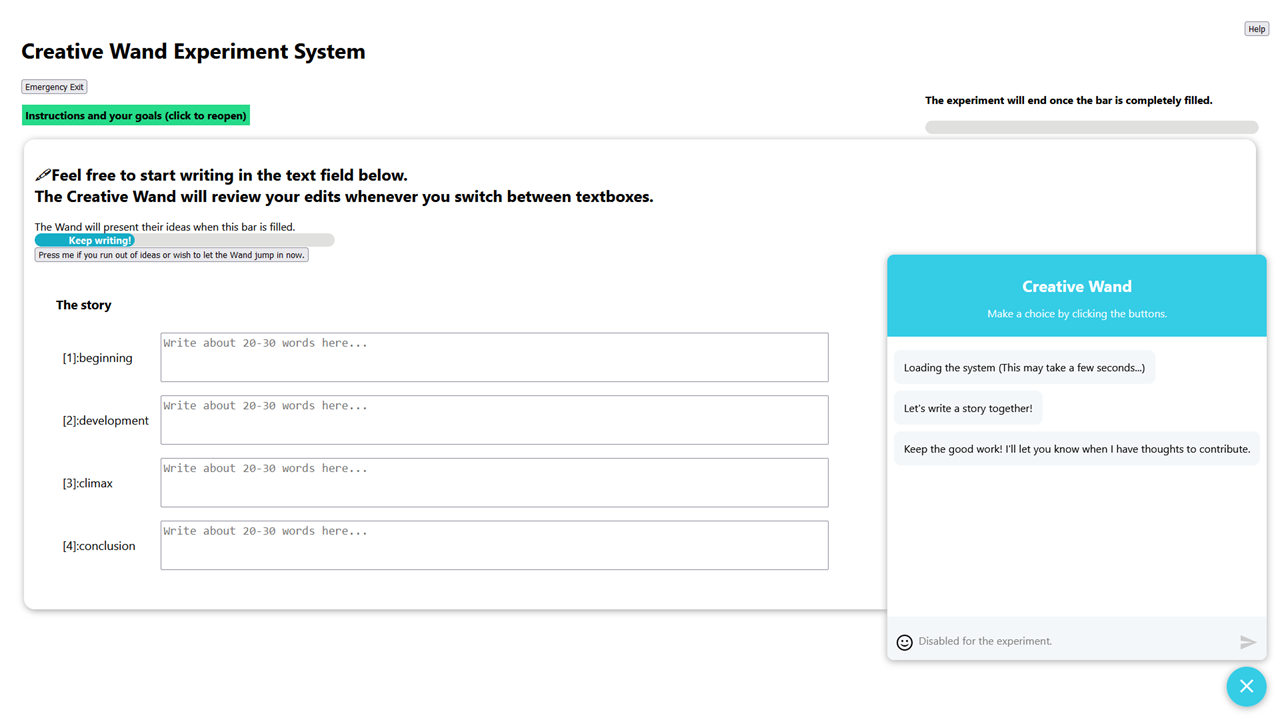}
    \caption{Our system in action. How should an AI learn and adapt to the creators in this Mixed-Initiative Co-Creative setting?}
    \label{fig:cw3-screenshot}
\end{figure*}

Recent advancements in Machine Learning (ML)--powered Artificial Intelligence (AI), such as large language models (LMs) \cite{openai_gpt-4_2023} and diffusion models \cite{dhariwal_diffusion_2021}, have made a new class of tools for Procedural Content Generation (PCG) available to game creators. 
The dominant contemporary way for the creators to control such {\em generative AI} models is via {\em prompting}---the issuing of textual instructions for the model to interpret and respond to~\cite{liu_pre-train_2021}.
That is, the user is tasked with the responsibility of issuing clear ``prompts'' to contextualize the AI system and make them aware of their intents. 
The AI is tasked to follow and fulfill the request strictly based on it.
If the system does not respond with an output that satisfies the creators' wants or needs, it is \textit{incumbent upon the creators} to modify the prompt and try again. 

The paradigm of human creators working with generative AI via prompting is just one of many theoretical ways for a human creator and an AI system to interact~\cite{lin_ontology_2023}.
There is evidence that prompting is not necessarily the best interaction paradigm; users indicate an appreciation for more varied ways of interacting with AI creative systems~\cite{lin_beyond_2023}.
Other configurations of human-AI collaboration creative systems are possible that promise to reduce cognitive load, frustration, and system abandonment \cite{sweller_cognitive_2011}, and make these systems more casual and enjoyable \cite{compton2015casual}. 
These include {\bf Mixed-Initiative (MI)} systems and {\bf Co-Creative (CC)} systems.
Mixed initiative (MI) systems are those in which both human and AI systems can initiate content changes.
Co-Creative (CC) systems are those in which both human and AI systems can contribute to content creation. 
In particular, MI-CC systems have been demonstrated in game design \cite{liapis_can_2016}, drawing \cite{davis_drawing_2015}, and storytelling \cite{alvarez_story_2022}, that benefits from both human and AI possessing the ability to take creative initiative.
While the broadest definition of co-creative systems might include any human creators working with a generative AI, the vast majority of them have not investigated the role of mixed-initiative, especially a more \textit{active} AI initiative.

At the heart of MI-CC systems is the question of whether and how the AI creative agent knows and understands (a)~the intentions and goals of the human creator and (2)~how the user wants to work with the AI system. 
These questions pose significant challenges, especially within domains critical to game designers utilizing AI, such as Computational Creativity and PCG.
In other domains, the goal may be provided to AI in advance, making it easier to identify opportunities to take the initiative with respect to contributing to a solution---the extreme of which is the AI system knowing the goal and solving the goal completely on its own.
When it comes to creating games, however, the human creators' intent is harder to articulate completely\cite{riedl_human-centered_2019}.
The human creator's goals are also non-stationary and may evolve during the creative process\cite{davis_enactive_2015, guzdial_co-creative_2018}.
The human creator might also have a \textit{preferred working style} that the agent should conform to in order to take the initiative while minimizing disruption.
Once we overcome these challenges, researchers have shown that such ambiguity and instability link to improved outcomes of the creative activity\cite{zenasni_creativity_2008}, thus benefiting the MI-CC interaction. 
%

In this paper, we examine Co-Creative systems in a mixed-initiative setting and study the dynamics of managing creative responsibility between human and AI initiatives.
We ask: \textbf{What influence does an AI agent's ability to actively adapt to creators' expectancy of creative responsibility in an MI-CC system have on creator experience and perception?}

In particular, we make the assumption that the AI agent is capable of working in the creative domain if given explicit prompts 
but is unaware of the human creator's preferences for distributing \textit{creative responsibility} between humans and the AI. 
We explore the usage of Reinforcement Learning (RL) methods in this setting and
demonstrate that \textbf{the creative responsibility learning challenge in MI-CC systems can be addressed by a multi-armed bandit (MAB) algorithm that observes feedback from users iteratively, updates its beliefs, and carries out its capabilities to facilitate the MI-CC collaboration.}
The learning is done online in real-time {\em during} the MI-CC process, and the human creator is not expected to have previous knowledge of the AI agent or time to pre-train it with regard to their collaboration style.

Working in the domain of {\em structured story co-creation},
we invite 39 participants to a human subject study.
We quantitatively measure the human creator's {\em perceived} learning performance of the agent and the overall level of satisfaction with the collaboration. 
We use the Creative Support Index (CSI)~\cite{cherry_quantifying_2014}
to study the implications of a learning and evolving AI agent. 
We also report on qualitative data collected from participants, using a grounded theory \cite{glaser_discovery_2017} approach in which we identify thematic patterns in users' subjective reports of their experiences.
This study revealed \textbf{a higher degree of participant recognition regarding the learning capabilities of our agent}, compared to the ablation, which in turn corresponded to \textbf{a significant increase in overall satisfaction with our agent.}  \footnote{The system used in the study will be open-sourced.}

\section{Background and Related Work}
The procedure of an MI-CC system learning its creative responsibilities can be described as a decision-making process, where the agent communicates with the human creator, gathers information, and chooses among its capabilities. 
This is not as straightforward as asking human creators to prompt AI agents because:
\begin{itemize}
    \item Just like the Cold Start problem experienced by AI agents lacking prior preferential knowledge from their creators \cite{bobadilla_collaborative_2012}, 
    human creators, even experts, may struggle to make inferences about the behavior of AI systems they initially face;
    \item The ability of human creators to effectively convey information to AI depends on their communication skills, which can be a significant obstacle even in human-to-human interactions \cite{grover2005shaping}.
    \item Enforcing this AI-centric method of input requires a profound mechanical understanding of the AI system
    from the human creators, where this knowledge does not necessarily intersect with their expertise.
    This marginalizes creators who do not possess the requisite expertise in utilizing AI.
\end{itemize}

For these reasons, relying solely on human creators for direct collaborative prompting, \textit{regardless of the capability of the AI models}, has its limitations, leading to efficiency, cognitive load, fairness, and equity issues.

Alternatively, a model can be built on human feedback, where the human creators instead provide ``good'' or ``bad'' feedback signals to indirectly improve the model.
These approaches differ from generative models that rely solely on user-defined goals, as the constraint of requiring such inputs can be alleviated.
When it comes to generating contents, this is the foundation of methods such as RL from human feedback \cite{ziegler_fine-tuning_2019}, that has proven to drastically improve the quality of generated text in state-of-the-art models such as GPT-4\cite{openai_gpt-4_2023}.
Yet, they are designed to exclusively optimize for a static, known-from-data objective.
They are not designed for online implementation where pre-training is not feasible, and the system lacks prior knowledge of new creators and needs to actively probe them.

To focus on the active probing challenge, we formalize it as a Multi-Armed Bandit (MAB) problem \cite{vermorel_multi-armed_2005} \textit{above generative abilities}, where an AI agent needs to \textit{actively choose} under uncertainty from their library of capabilities based on their understanding of their human creator teammate, to minimize total regret and maximize rewards from their teammate.
Multi-Armed Bandit systems have been employed in the context of resolving how to make progress in an interactive creative experience.
Koch et al. \cite{koch_may_2019} discussed a design ideation framework that suggests images that a designer may like by exploring and exploiting in the image embedding space with a variant of MAB; 
Gallotta et al. \cite{gallotta_preference-learning_2023} applied MAB in the context of generating ``in-game spaceships'' by enabling creator-guided latent space walk in the feature embedding space representing such spaceships.
These works focused on a single type of action in the content space, and concentrated on expanding the generative space of such content; 
Lin et al. \cite{lin_creative_2022,lin_beyond_2023} explored instead the \textit{action space}, characterized as types of Communications representing information exchange between human and AI used in the co-creative process;
As to the idea of switching between different high-level actions beyond the content level, 
Building a model of the user has been proven to help in a CC setting; specifically in the domain of storytelling.
Yu et al. \cite{yu_data-driven_2013} demonstrated its potential to generate stories that bring ``an enjoyable experience for the players'';
Gray et al. \cite{gray_player_2020,gray_multiplayer_2021} further demonstrated how MAB agents help to capture this player model.
Vinogradov et al. \cite{vinogradov_using_2022} showcased a framework where the agent explores the creators' ``player'' model vigorously by directly generating ``distractions'', objects designed to probe into players' preference instead of providing utilities in finishing a certain task;
They proposed using MAB for this task for its promises in ``balancing the act of gathering information about the payout associated with each arm (exploration) and maximizing reward given the current known information (exploitation)'', 
dynamically updating the model in the process towards assigning tasks that the players feel more interested in tackling.
They inspire our method, as its approach of adding distractions is well comparable to the agent carrying out its initiative while directly changing the creative content. 

\section{Study Design}
\label{paper_cw3_mab}


In this section, we present the study we designed to examine the AI agent we created that adapts to creators' expectancy of creative responsibility.
We seek to determine how this changes the \textbf{perception of the creators toward the AI} and \textbf{the creative experience} the system supplies to the human creators.

\subsection{Task Setup}
\paragraph{The Delegation Setup.}
For the experiments, we spotlight a specific but generalizable collaborative setup: \textit{Learning a delegation}.
In this setup, both parties would take a subset (or entirety, if preferred) of responsibilities in an MI-CC activity towards the common goal.
The human creator concentrates on specific parts of the creative task while not losing control of the other parts; the AI agent needs to strategically shift its focus towards the parts that the human creator is not focusing on and actively determine how to make improvements.
Furthermore, as these interactions are not without cost, such as creators' cognitive load, it is also important to minimize such costs towards learning these responsibilities.
We denote the expected and delegated responsibility that the AI agent needs to learn during the interaction preferred \textbf{work style} for a particular human creator.


\paragraph{Domain chosen: Storytelling.}
Given the established research foundation within story generation, its high relevance to game development, and its inherent complexity with regard to PCG, we selected story generation as a proving ground for our proposed method. 
The expertise of the team and advancements in open-source Large LMs readily available to us facilitated implementation;
This allowed us to focus on the human factors of the MI-CC experience and the AI agent itself.

For our experimental system, We used \textit{Llama2-13b-chat} \cite{touvron_llama_2023} as the LM, readily available at the time of the study while very responsive for the interactive experience.

\subsection{Experimental AI System overview}
\label{section:cw3-system-overview}
We now describe the AI system we built for the purpose of the study.
The experimental system is based on the Creative Wand framework \cite{lin_creative_2022}, containing the following four components: 

\subsubsection{Creative Context}
The Creative Context is the abstraction of generative models for this system.

We tuned a model to generate a story containing four components inspired by the Narrative Arc theory.
We sketched out the framework of a story with four components, which are the beginning, development or rising action, climax, and conclusion.
Each component is designed to hold multiple sentences.
Both the human participant and the AI are instructed to write about 20 to 30 words per component, and the target length of the whole story is around 100 words.


Once we set up the model, it will take requests from \textit{Communications}.


\subsubsection{Communications}
\label{paper-subsection-comm}
Communications describes the interactions between the human creators and the AI;
they also double as 
the \textbf{capabilities} the AI agent possess.
Three communications are used in this study, a minimalistic but complete set for the task:
\begin{itemize}
    \item (Re)write the beginning and development;
    \item (Re)write the climax and conclusion;
    \item Write a review of the story, one sentence positive, one negative, and one suggestion for improvements.
\end{itemize}
As we focus on how the agents would learn, a smaller set of communications is chosen, allowing the participants to focus on research questions about the creative experience while minimizing the cognitive load of learning the system.

Each of the communications includes a prompt to the LM describing the responsibilities.
See Appendix \ref{appendix:prompts} for details.

\subsubsection{Experience Manager and Frontend}
These two modules manage the interactive experience and workflow.

\begin{figure*}
    \centering
    \includegraphics[width=0.9\textwidth]{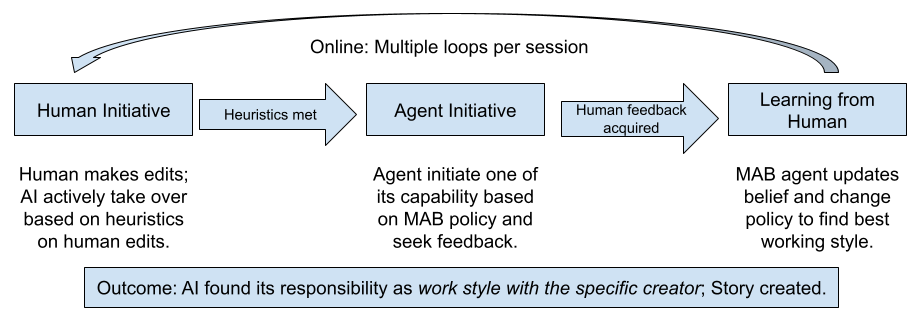}
    \caption{One round of interaction of our experimental system. Each participant will experience multiple turns per session.}
    \label{fig:cw3-diagram_phases}
\end{figure*}

We implemented a Finite State Machine to manage the experience.
Figure \ref{fig:cw3-diagram_phases} shows the states with the overall flow of interaction each participant experiences in one experiment session.
One session of the MI-CC experience is separated into multiple ``turns'', where both parties iteratively improve the story, sharing the same text fields in the editing process.
The participants are not directly notified of the internal states of the system.

\paragraph{Human Initiative.}
During this phase, human creators contribute to the story by making edits in any of the four text fields.
This phase ends when the agent decides to take the initiative.
We implemented a point-based heuristic based on  pilot studies:
the agent would assign points for changes it observes, and will take initiative whenever enough points are accumulated, signifying substantial edits from the human creators, in the following criteria:
\begin{itemize}
    \item Each new character would add 5 points;
    \item Each time the human creator switches between fields after any changes, 100 points are added;
    \item Whenever the human creator leaves a text field with 200 points accumulated (roughly one full sentence or two minor changes), the agent will take the initiative by locking the editing interface and resetting the counter.
\end{itemize}
This heuristic provides two advantages compared to other ways this decision can be made: 
First, this heuristic is computationally fast and enables responsive interactions;
Second, it additionally provides visualization for the users. 
As shown in Figure \ref{fig:cw3-screenshot}, we present this right above the text boxes for the stories, with a text hint and a progress bar representing the ideation process of the agent.
We additionally provided a ``skip'' function that forces agent initiative.

\paragraph{Agent Initiative.}
In this phase, the agent decides which capability best fosters the collaborative experience and carries out the corresponding Communication.
We built a Multi-Armed Bandit-based agent that is responsible for choosing which Communication to invoke, with Thompson Sampling as the chosen algorithm for the experimental system, within the AI agent.
Formally, an agent $A$ interacts with a set of $K$ arms $a_1 \cdots a_{k}$, each of which is associated with Communication and underlying capabilities and an unknown reward distribution. 
Whenever an arm is pulled, the agent seeks feedback from the human creator on the initiative, which is treated as a reward signal. (See next paragraph.) 
The goal of the agent is to maximize the total reward obtained by repeatedly pulling arms during the session.
See subsection \ref{appendix:mab} for more details on the design choices of the MAB agent.
Once an arm is pulled, the agent executes a Communication, interacts with the user, and updates the story as needed.

\paragraph{Learning from human.}
The system would ask about (Action Feedback) the way they just worked and (Content Feedback) the updates and content changes.
The participants choose between ``Good'' (Reward of 1) and ``Bad'' (Reward of 0).
``Bad'' feedback on generated text leads to a reversion to the original content, though it is not used to improve the LM in any way.

A weighted mean is employed to integrate both types of feedback into a singular reward signal.
For the study, a weight hyperparameter of 80\% is applied to the Action Feedback and 20\% to the Content Feedback.
This prioritizes learning action-level responsibilities rather than the preference for LM-generated text, in which the full system and the baseline share implementation.
This reward signal is then used to train the agent.

For this experiment, an MAB agent with Thompson Sampling is used in the experimental system.
See \ref{appendix:mab} for a discussion and experiments related to this choice.

Once the learning process is complete, ``human initiative'' starts again.
To maintain user engagement, text responses are morphed each time to avoid repetitiveness, while contextual hints are also strategically provided throughout the experience.
Figure \ref{fig:cw3-screenshot} shows the user interface.

\subsection{Study Methodology}
\label{section:cw3-study-methodology}

\begin{figure*}
    \centering
    \includegraphics[width=0.9\textwidth]{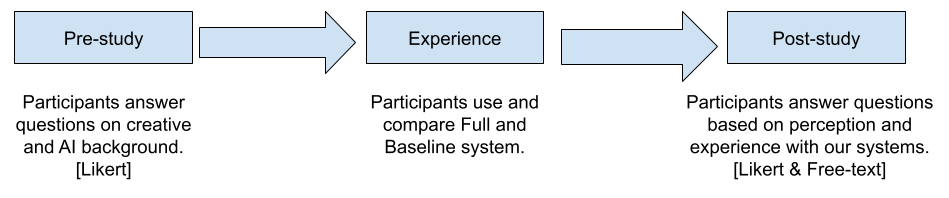}
    \caption{Participants' experience during the study.}
    \label{fig:cw3_diagram_study}
\end{figure*}
To study the perception of human creators towards MI-CC systems equipped with these learning capabilities, we conducted a study summarized in figure \ref{fig:cw3_diagram_study} on the AI system.

We compare our system, the ``Full'' system, with an ablation named ``baseline''.
The ``baseline'' ablation does not learn. 
It chooses each of the 3 Communications with a $1/3$ probability at all times and provides only a reverting option when ``asking for feedback''.
These systems are codenamed ``Echo Wand'' and ``Harmony Wand'' respectively, not to reveal the details of the systems to the participants during the study.


We recruited 39 United States participants \footnote{Only counting participants who finished the whole study with valid sessions and responses.} on Prolific\footnote{\url{prolific.co}} with adequate English proficiency.
Each experiment session lasted for approximately 40 minutes, and we paid the participants \$15 per hour for perfect completion of the study.

\paragraph{Pre-study.} Before the experience, participants answer four 5-point Likert-scale questions on (Q1) Expertise in Computer-Assisted Designing (CAD), (Q2) Expertise in writing stories, (Q3) Frequency using AI, and (Q4) Understanding of AI. \footnote{See Appendix \ref{appendix:questionnaire} for the full question text.}

We then present instructions to familiarize the participants with our systems by providing annotated screenshots of the interface, which is a copy of \ref{fig:cw3-screenshot}, but with additional numeric overlays,  descriptions of components, and a brief introduction to the workflow of co-creating a story.
 
They are then assigned the delegation task to focus on writing the beginning and the development of the story while leaving the other parts of the story to AI as much as possible. 
They are also made aware that the AI does not know this setup in advance.

\paragraph{Experience.}
Participants are assigned to interact with the full system and the baseline ablation, presented in random ordering, counter-balanced.
They are given 10 turns per each of the 2 sessions. 

\paragraph{Post-study.}
After participants finished two sessions using our system, they were asked about the process they had just experienced.
inspired by Creative Support Index (CSI)~\cite{cherry_quantifying_2014} used in the previous studies,
We ask questions based on dimensions related to the creative support perception and overall collaborative experience, grouped to facilitate richer responses from the participants while maintaining their engagement in the survey.

Specifically, we asked which system(s), are (Q5, Learning, Collaboration) learning to collaborate, (Q6, Enjoyment, Immersion) more capable and easy to work with, (Q7, Expressiveness, Exploration, Results worth effort) enabling better stories;
For Q5 through Q7, participants can choose either system, both systems, or neither to be chosen, leading to a potential total exceeding 100\%.
We ask one final question (Q8) on which system will they recommend more, framed in a win-draw-lose format.

Although these questions are presented in the same order for all participants, the order of the options is randomized to reduce bias towards any system.
All questions are followed by an open-text question prepared to collect justifications from the participants.

\section{Results}
\subsection{Creative background}

\begin{table*}[!ht]
    \centering
    \caption{Creative background of the participants. 1 = Most Negative, 5 = Most Positive.}
    \textbf{}
    
    \begin{tabular}{|c|c|c|c|c|c|c|c|}
        \hline
        \textbf{Q (See Appendix \ref{appendix:questionnaire} for full questions)} & 1 & 2 & 3 & 4 & 5 & Average & Median \\ \hline
        \textbf{Q1: CAD skills} & 1 & 1 & 2 & 19 & 16 & 4.23 & 4 \\ \hline
        \textbf{Q2: Writing skills} & 1 & 0 & 7 & 20 & 11 & 4.03 & 4 \\ \hline
        \textbf{Q3: Frequency of using AI} & 0 & 0 & 16 & 11 & 12 & 3.90 & 4 \\ \hline
        \textbf{Q4: Understanding of AI Tech.} & 0 & 5 & 14 & 19 & 1 & 3.41 & 4 \\ \hline
    \end{tabular}
    \label{table:cw3-participants}
\end{table*}


Table \ref{table:cw3-participants} shows a summary of the creative backgrounds of the participants.
Although a median of 4 on all questions implies that participants are familiar with the recent advancement of AI, 
when specifically asking whether they can build one, only 1 participant answered ``yes'' (5 in Q4), meaning that most of the participants do not have a technical background.

However, comparing to 26\% reported in \cite{lin_beyond_2023}, we observed 87\% of the participants at least being ``somewhat familiar'' (3+) with recent AI technologies, and 51\% being ``familiar'' (4+);
The experience of using commercially available Large LM-based agents may have a profound effect on how participants, in general, would collaborate with AI systems.

\subsection{Quantitative Results}
We commence by presenting the quantitative results of the study through the choices made by the participants in the multiple-choice questions.


When asked which system(s) learned to collaborate with them under the delegation arrangement (Q5), the ``Full'' system is chosen 69\% ($n=39$) of the times, compared to 51\% for the baseline ($p<0.018$, under a binomial test where $H_0 :=$ no observable difference in distribution; The same for all p-values in this section).
We clearly see \textbf{the ``Full'' system with learning capabilities enabled being perceived significantly better at learning the delegation than the baseline}, demonstrating the effectiveness of the MAB-based model \textbf{From the human creator perspective} learning from their feedback.

When asked which system to recommend, this trend also persists:
Our system is preferred (wins) 43.6\% of the time, versus 20.5\% (loses) for the baseline ($p<0.001$); 35.9\% of the participants do not have a preference (draw).
The ``Full'' system is only different from the baseline system with the learning capabilities and corresponding frontend elements, yet we see \textbf{a statistically significant improvement in preference towards our ``Full'' system}, illustrating the potential of our method in enhancing MI-CC experience and making such system better for human creators.


When it comes to which system(s) gave a good story (Q7), 72\% of the participants agree that the ``Full'' system made a good story, while 69\% selected the baseline system ($p>0.05$).
We were unable to statistically determine whether an agent learning the delegation would produce a better story;
This is expected, 
We focused on studying the sharing of responsibilities and enforced a delegation setting.
In an actual MI-CC experience, without such a prior, A human creator would utilize the agent's learning capability to promote their strengths and discourage their weaknesses, and an improvement in perceived performance is more likely to be observed in that setting.

Finally, when queried about the collaboration itself (Q6), 62\% of the participants think the ``Full'' system is capable and made the collaboration easy, while 56\% voted for the baseline system (p>0.05).
We also were unable to statistically determine whether the ``Full'' system is more enjoyable and immersive.
Although the difference between the ``Full'' system and the baseline is substantial enough both implementation-wise and towards the perception of learning, 
from the angle of the user interface, the only difference is 10 additional questions from the ``Full'' system per session.
Previously, Larsson et al. \cite{larsson_how_2023} reported that ``there was a clear trend that the visual ... was rather important to the subject's relationship towards the MI-CC.'' while these ``relationships'' are directly linked to creators' perception of immersion of the experience;
Ehsan et al. \cite{ehsan2021explainable} additionally pointed out that even when an AI system presents the same underlying information, how it is presented influences the perceptions of human users.
We may have observed this effect from a different angle, where a lack of differences in presentation may have caused the indifference of the participants.
To that end, the difference between the two systems on these creative support dimensions may be too minor \textbf{when it comes to how they are presented visually};
The effect of \textit{user interface} used to present the results in an MI-CC system is out of the scope of this work, though these findings illuminated a potential path for future research. 

\subsection{Qualitative Results}
We now show the results from the open-ended questions following each multiple-choice question.
Open-ended justifications participants provided for each of the four questions are evaluated with thematic analysis~\cite{aronson_pragmatic_1994}, based on grounded theory~\cite{glaser_discovery_2017}.
Taking an inductive approach, we started the process with an open-coding scheme and iteratively produced in-vivo codes (generating codes directly from the data).
Next, we analyzed the data using axial codes, which involves finding relationships between the open codes and clustering them into different emergent themes. 
Through an iterative process performed until consensus was reached, we share the most salient themes that emerged from axial codes.

\paragraph{A MI-CC system that understands the intents of the human creators and follows them by learning is overall favored and collaborates well with the creators.}
Participants demonstrated their observation of the learning capabilities of the ``full'' system, identifying them as ``better about learning that I specifically wanted help with'' (P34)
and ``listened to my feedback.''(P39).
In comparison, the baseline system is identified as ``did less of the work ... did not necessarily learn what its role was expected to be'' (P19).
this resulted in a preference for the Full system for P32, as the Full system is quoted as a ``more useful helper".
This aligns with the quantitative observations.

\paragraph{Good content suggestions may give people the feeling that the system is learning how to collaborate with them, regardless of how AI is actually doing so.}
Despite specifically asking participants to discuss whether the agent has ``learned to collaborate with you under that arrangement'' (Q5),
Participants are also rating the system based on the generated content:
\begin{quote}
   (P25, emphasis asked)
    This one learned from me because \textbf{it was able to build off of my original foundation of my story} that I typed. 
\end{quote}
P18, who rated their familiarity with AI as Familiar (4 out of 5) and AI usage as ``Always / as much as possible'' (5 out of 5), wrote that the ``Full'' system is learning from them:
\begin{quote}
    I could see Echo Wand adding more detail and building out more creatively than with Harmony Wand. 
\end{quote}
This participant is familiar with recent generative AI and mentions ``adding details'' and ``building,'' which are traits that these AI are optimized for.
As both the ``Full'' and the baseline use the same underlying generative AI capabilities, P18 could not distinguish between the ``improvements'' on generated contents and the performance of the MAB-based agent.
The apparent improvements of generated stories may result from a wide range of reasons, such as participants providing different input and LM sampled differently, unrelated to both the underlying LM and the learner, creating noises in the perception of participants.  



\paragraph{Diversity is also important, it may not be the best strategy for a learning agent to pick the “best options”, and sometimes the agent may want to intentionally surprise their teammates.}

P23 was impressed by the range of capabilities both agents possess, seeing ``They were both impressive, being able to take my story and to word it better, or even add things to change it to make it better''.
When asked about the generated story, P39 mentioned that `` Both of them gave bad stories.'' and  ``I need much more control and options''. Curiously, this is the same participant that enjoyed the agent that ``listened to my feedback.''.
P36 preferred the baseline system that executes random actions:
\begin{quote}
    I did all of the work with Echo, despite my best efforts to get it to collaborate with me. Harmony had much more interesting suggestions and rightfully pointed out when a section became too dense. It balanced the second two sections to match my intro and build up, unlike Echo who almost refused to work on them.
\end{quote}
For this study, we assigned delegation tasks to the participants.
This is only a subset of possible responsibilities that the AI agent can take and the human creators may expect. 
Lin et al. \cite{lin_beyond_2023} have shown that a system with more coverage of the design space, providing more diversified options, is preferred.
Our study design, which is more focused on studying the learning process, limited the variety of capabilities the agent may perform.
To that end, once such an MI-CC system is put into use beyond research, it is necessary to diversify both the capability pool and the process of the AI agent choosing them, potentially providing surprise and unpredictability to further inspire the users.


\paragraph{Creator control is important, and creators may want their ideas to be included even when AI can provide better candidates.}
Beyond the need for control mentioned by P39,  
P28 mentioned that they were impressed by the capabilities of both systems in ``finish the story \textbf{that I started with.}'' (Emphasis added).
P27 mentioned further on their justification:
\begin{quote}
    ... I was in control of the final text to accept changes or not, or to make my own.
\end{quote}
In a system involving a creator who wishes to create content to their liking, it is expected that the creator wishes to solicit as much control as possible.
However, if the AI agent does not have any final say on the contents, should we expect it to take any creative responsibilities?
Although we acknowledge that this is more of a philosophical question, way out of the scope of our work, what if the agent would understand what their counterpart is actually seeking and use this information to determine what contribution they should stick to by understanding what \textit{human creators are thinking}?

\section{Discussions}

Distilling from these findings ranging from the perception of collaboration, good writing skills, diversity in capabilities, and creators' need for control, a common implication surfaces: \textbf{Getting the mental model of the creators right, the system will succeed; Getting it wrong, failure cases would surface.}
A mental model is described by Kieras et al. \cite{kieras_role_1984} as `` understanding ... that describes the internal mechanism`` of the system a human is operating; 
Leslie et al. \cite{leslie_core_2004} further point out that a theory of mind is a mechanism that human expresses naturally, towards an understanding of \textit{thinking}, in our context, their teammate AI. 
The success of our ``Full'' system of learning comes from its ability to learn a model of how the creators wish to collaborate with them; 
Although we didn't formalize this as learning a mental model, the reward given from a teammate can be otherwise treated as a \textit{reward for correctly understanding the mental model of the human creator}.
The need for diversified responses and more respect to control signals users imposed also fall into this paradigm, but beyond;
Understanding how these reward signals should be used to determine initiative from the agents beyond ``picking the best'', and how to capture hints of additional types of actions or capabilities needed can greatly improve collaborations with MI-CC systems.
This falls into the subfield of ``novelty detection and adaptation'' \cite{balloch_novgrid_2022} situated in RL, which is known to be challenging, if solvable at all with ML methods, as ML models can only rely on their extrapolation capabilities towards the ``unknowns'', that may not hold for all novelties;
This will be a rewarding pathway towards better MI-CC systems, if not AI agents overall.

We start to see a consistent narrative: creators are interpreting the capabilities of our AI agent learning as an attempt the AI agent made to learn a mental model of themselves;
Because our agent determines which Communication to use and the effect of it on the contents being collaborated on, 
We observe the participants treating proper learning of Communication choices (expected) and the content generated (emerging) as both evidence that the agent is learning from them and traits leading to their preferences towards these systems.
This also, to some extent, explains the placebo effect we observe on the baseline system: 
around half of the participants believe that the baseline system is learning from them, significantly more than 0, despite the baseline system only making decisions randomly.
In this controlled comparative study, to avoid a bias towards either of the systems, we intentionally did not disclose any difference between the ``full'' system and the baseline.
This perception may have arisen from the capability of our agent to generate part of stories that follow the context that the participants provided.
Although we acknowledge that these factors are hard to decouple, this finding also hints at the potential of our methods in understanding the human creator \textit{holisticly}.
Upol et al. \cite{ehsan2021explainable} 
pointed out that the background of human users determines their cognitive heuristics, which plays a role in their expectations \textit{beyond what the designer of the systems expected in the first place}.
They also realized that if not treated carefully, AI systems can actually introduce such placebo effects, as a pitfall \cite{ehsan_explainability_2021}, by misleading the human users into appreciating their trustworthiness and power, without the development of underlying AI capabilities.
Standing on these findings, A promising direction of research is to carefully identify the effect of \textit{expectations} of both parties involved in the MI-CC process, and how they dynamically change during the collaboration.

\section{Conclusions}
In this paper, we showcased how an MI-CC system is capable of listening to human feedback and improving itself towards a better understanding of how it should collaborate with human creators in a storytelling domain.
Inviting 39 participants and comparing two such systems with and without these learning capabilities,
we found that this capability was well recognized by the participants and led to better satisfaction overall.
To this end, we further encourage the designers of MI-CC systems to pay attention to both the human creators and the AI agent, study how each party should, or is already, \textbf{adapting to and creating mental models of their counterpart}, based on their creative roles taken, their previous experience, and capabilities, and most importantly, \textbf{the wishes of the human creators}.


\section{Appendices}



\appendix
\section{Choosing a MAB algorithm}
\label{appendix:mab}

\begin{figure}
    \centering
    \includegraphics[width=\linewidth]{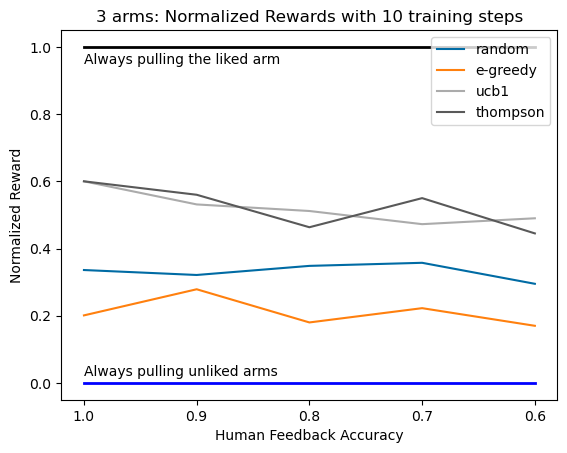}
    \caption{Oracle experiment results on MAB algorithms of the agents performing on various feedback accuracy levels. Upper Bound performance, where the liked arm is always pulled, and the Lower Bound, where one not-liked arm is always pulled, is also presented for reference.}
    \label{fig:cw3_oracles}
\end{figure}


In this section, we provide more information on the design choice of  the MAB agent.
Following results from Vinogradov et al. \cite{vinogradov_using_2022},
we looked into three representative MAB algorithms: $\epsilon$-greedy, UCB1 and Thompson Sampling.

$\epsilon$-greedy~\cite{sutton_reinforcement_2018}, widely used in RL, works on a simple principle:
The agent has probability $\epsilon$ (a hyperparameter) to choose a random action (explore) instead of performing the best action from its policy (exploit).

UCB1, or Upper Confidence Bound 1 \cite{agrawal_sample_1995} instead takes a more deterministic approach:
This algorithm calculates an ``Upper Confidence Bound'' for each arm, considering both the current running average of the rewards and the uncertainty due to lack of sampling:
\begin{equation}
a = argmax(\bar{x}_a+\sqrt{2 \log t / n_a})
\end{equation}
where $\bar{x}_a$ represents the average reward received from arm $a$, $n_a$ represents number of times arm $a$ was pulled, and $t$ the total number of times all arms are pulled.
This makes UCB1 aware of the uncertainty of the rewards from each arm when the agent makes its decisions.
Although probability distributions are used to calculate these bounds, this algorithm does not sample at all and provides a deterministic choice for a given system state.

Finally, Thompson Sampling is a robust Bayesian approach first introduced by Thompson \cite{thompson1933likelihood}.
It maintains a probability distribution over the possible values of each arm's reward and uses this distribution to make decisions.
To determine which arm to pull, it draws samples from a Beta ($\mathcal{B}$) distribution of the number of successes and failures for each arm, choosing the sample with the maximum probability, while seeking a reward between 0 and 1:
\begin{equation}
    a = argmax(\mathcal{B}(\alpha_a,\beta_a))
\end{equation}
$\alpha_a$ increases by the reward received,
and $\beta_a$ increased by 1 minus the reward received. 
Initially, both $\alpha$ and $\beta$ for each arm are set to 1 to establish a uniform prior distribution.
Thompson sampling is designed to effortlessly transition from primarily exploring in the initial stages to a more exploitation-oriented strategy as it acquires more information. 


We carried out an Oracle-based experiment to determine the MAB algorithm of choice for the study. 
Using an oracle, which simulates a human creator interacting with the system, gives us total control of their behaviour.
We measure the performance of the agents at various levels of \textit{human feedback accuracy}, to seek an agent that generally performs well on all accuracy levels so that it serves a wider variety of human creators well.

We study four different agents and baselines:
$\epsilon$-greedy, UCB1, Thompson Sampling, and 
Random Baseline, where a universally random arm is chosen each time.
We give the agents 3 arms to pull, where one is ``liked'' and two others are ``unliked''.
Each arm would give either a reward of 1 if liked or 0 otherwise when pulled, by the oracle;
We define human feedback accuracy as the probability of the oracle giving a reward of 1 on pulling the ``liked'' arm and a 0 on pulling the ``not liked'' arm.
As this value gets lower, closer to 50\%, the simulated oracle becomes less clear on which arm it liked and becomes a less efficient feedback provider.
We simulated 5 levels of this accuracy, from 60\% to 100\% with equal intervals.

$\epsilon$-greedy is highly sensitive to its $\epsilon$ parameter chosen, and we report with the best performing $\epsilon$-greedy agent in the with $\epsilon = 0.2$.
We report the ``normalized rewards'', which is the agent's reward relative to the theoretical maximum of always choosing the ``liked'' arm.
We repeat each experiment condition 100 times and report the mean normalized rewards after 10 steps to simulate a scenario where the MI-CC agent has to quickly learn from their human counterparts, similar to our actual study.

Figure \ref{fig:cw3_oracles} summarizes the results from the Oracle experiments.
As we only gave these agents 10 steps to learn the arms, 
the agent may not have yet converged.
This is expected in a quick-learning scenario.
$\epsilon$-greedy performed poorly, even worse than the random baseline, likely due to its inability to quickly change focus between exploration and exploitation;
UCB1 and Thompson perform at similar levels, demonstrating their capabilities to calculate an upper-bound reward and use it in their decision-making process.

Although UCB1 and Thompson performed similarly, Thompson Sampling is preferred because of its sampling behavior.
UCB1 schedules its exploration over a very long session in a deterministic way (exploring once after exploiting $n$ times).
As we aim for quick learning and adaptation, without sampling, UCB1 risks showing ``stubbornness'' to a suboptimal arm without any \textit{probability} to unstuck itself, a behavior that is less preferred from an MI-CC perspective.
Thompson Sampling, on the other side, exhibits its capability to \textit{dynamically change its exploration aggressiveness based on previous observations}, while \textit{using a Bayesian prior instead of greedy sampling}, both benefiting its application in our experiment MI-CC setup.
This results in both an effectively dynamic ``epsilon'' compared to epsilon-greedy and some randomness instead of being fully greedy per each step, compared to UCB1.


We chose Thompson Sampling as the MAB algorithm used in the experimental system.

\section{Questionnaires used in the study}
\label{appendix:questionnaire}

\paragraph{Pre-study.} Four 5-point Likert scale questions are asked:
\begin{itemize}
    \item Q1: Do you agree that you are familiar with the process of creating content, such as writing articles, drawing pictures or creating a video game stage, using a computer? (Strongly Disagree $\rightarrow$ Strongly Agree)
    \item Q2: Do you agree that you are good at writing or telling a story, either real or fictional? (Strongly Disagree / Never attempted in the past 5 years $\rightarrow$ Strongly Agree)
    \item Q3: How frequently do you use or interface with artificial intelligence? For example, using map services to find a route to your destination, playing a game with a computer-controlled character, or using a chatbot. (Never used $\rightarrow$ Always / For as many things as possible)
    \item Q4: How much understanding do you have of the recent developments in Artificial Intelligence technologies? (Very unfamiliar  $\rightarrow$ Very familiar / I can build one)
\end{itemize}

\paragraph{Post-study.} 
Four questions are asked regarding the systems they used during the study.
\begin{itemize}
    \item Q5-\textit{(Learning, Collaboration)} You were assigned a specific way to collaborate with the assistant Wands, and the assistant is not informed of this arrangement in advance. 
    Which assistant wand learned to collaborate with you under that arrangement?
    If you have chosen at least one of the assistant wands, how did you know they learned from you?
    \item Q6-\textit{(Enjoyment, Immersion)} Which assistant wand is more capable and made the collaboration easy for you?
    If you have chosen at least one of the assistant wands, how did the assistant(s) impress you with their capabilities?
    \item Q7-\textit{(Expressiveness, Exploration, Results worth effort)} With these assistant wands, which collaborative experience ended up in a good story?
    If you have chosen at least one of the assistant wands, What do you think helped? If you chose neither, what went wrong?
    \item Q8-Lastly, which assistant wand would you recommend more to a friend or a colleague story writer?
    Please let us know if you have any other message or comment to share.
\end{itemize}
For Q5 to Q7, Participants may select one, both, or neither system; 
For Q8, as it is a comparative question, the option of "neither" is not available.
All questions are followed by an open-text question prepared to collect justifications from the participants.

\section{Prompting details}
\label{appendix:prompts}
Prompts for Communications start with
\begin{quote}
    ``You are an AI writing assistant, collaborating with a human on the task of writing a story.You are very concise, and answer only what is absolutely necessary, without any explanations or introductions.You make sure that all your answers are surrounded by an underscore, such as \_My answer\_ .''
\end{quote}
and are followed by a few examples of the tasks, along with the constraints, formed in a question-answering format;
The final question does not come with an answer, and the continuation is treated as the response.

\bibliography{references}

\end{document}